\documentclass{bmcart}

\usepackage{amsthm,amsmath}
\usepackage[breaklinks=true,colorlinks=true,linkcolor=blue,urlcolor=blue,citecolor=blue]{hyperref}
\usepackage[utf8]{inputenc} 
\usepackage{eurosym}
\usepackage{colortbl}
\usepackage{multirow}
\usepackage{graphicx}
\usepackage{nicefrac}
\usepackage{textcomp}
\usepackage{epstopdf}

\usepackage{graphicx}

\usepackage{subfigure}
\usepackage{array}
\usepackage{tabu}
\usepackage{braket}
\usepackage{multirow}
\usepackage{color, colortbl}

\newcolumntype{L}[1]{>{\raggedright\let\newline\\\arraybackslash\hspace{0pt}}m{#1}}
\newcolumntype{C}[1]{>{\centering\let\newline\\\arraybackslash\hspace{0pt}}m{#1}}
\newcolumntype{R}[1]{>{\raggedleft\let\newline\\\arraybackslash\hspace{0pt}}m{#1}}

\startlocaldefs
\endlocaldefs

\begin{document}

\begin{frontmatter}

\begin{fmbox}
\dochead{Research}

\title{Quantum Communication Uplink to a 3U CubeSat: Feasibility \& Design}


\author[
   noteref={n1},
   addressref={aff1,aff2},                     
   email={sebastian.neumann@univie.ac.at},   
]{\inits{SPN}\fnm{Sebastian Philipp} \snm{Neumann}}
\author[
    noteref={n1},
   addressref={aff1,aff2},
   email={siddarth.koduru.joshi@oeaw.ac.at}
]{\inits{SKJ}\fnm{Siddarth Koduru} \snm{Joshi}}
\author[
   addressref={aff1,aff2},
   email={matthias.fink@oeaw.ac.at}
]{\inits{MF}\fnm{Matthias} \snm{Fink}}
\author[
   addressref={aff1,aff2},
   email={thomas.scheidl@oeaw.ac.at}
]{\inits{TS}\fnm{Thomas} \snm{Scheidl}}
\author[
   addressref={aff1},
   email={roland.blach@oeaw.ac.at}
]{\inits{RB}\fnm{Roland} \snm{Blach}}
\author[
   addressref={aff3},
   email={Carsten.Scharlemann@fhwn.ac.at}
]{\inits{CS}\fnm{Carsten} \snm{Scharlemann}}
\author[
   addressref={aff3},
   email={sameh.abouagaga@fhwn.ac.at}
]{\inits{SA}\fnm{Sameh} \snm{Abouagaga}}
\author[
   addressref={aff3},
   email={daanish.bambery@fhwn.ac.at}
]{\inits{DB}\fnm{Daanish} \snm{Bambery}}
\author[
   addressref={aff4},
   email={erik.kerstel@univ-grenoble-alpes.fr}
]{\inits{EK}\fnm{Erik} \snm{Kerstel}}
\author[
   addressref={aff4},
   email={mathieu.barthelemy@univ-grenoble-alpes.fr}
]{\inits{MB}\fnm{Mathieu} \snm{Barthelemy}}
\author[
   addressref={aff1,aff2},
   corref={aff1},
   email={rupert.ursin@oeaw.ac.at}
]{\inits{RU}\fnm{Rupert} \snm{Ursin}}


\address[id=aff1]{
  \orgname{Institute for Quantum Optics and Quantum Information Vienna}, 
  \street{Boltzmanngasse 3},                     %
  \postcode{1090}                                
  \city{Vienna},                              
  \cny{Austria}                                    
}
\address[id=aff2]{%
  \orgname{Vienna Center for Quantum Science and Technology},
  \city{Vienna},
  \cny{Austria}
}
\address[id=aff3]{%
  \orgname{University of Applied Sciences Wiener Neustadt},
  \street{Johannes Gutenberg-Strasse 3},
  \postcode{2700}
  \city{Wiener Neustadt},
  \cny{Austria}
}
\address[id=aff4]{%
  \orgname{University Grenoble Alpes, Laboratoire Interdisciplinaire de Physique},
  \street{140 Rue de la Physique},
  \postcode{38400}
  \city{Saint-Martin-d'H\`eres},
  \cny{France}
}


\begin{artnotes}
\note[id=n1]{Equal contributor} 
\end{artnotes}

\end{fmbox}


\begin{abstractbox}

\begin{abstract} 
Satellites are the most efficient way to achieve global scale quantum communication (Q.Com) because unavoidable losses restrict fiber based Q.Com to a few hundred kilometers. We demonstrate the feasibility of establishing a Q.Com uplink with a tiny 3U CubeSat (measuring just 10$\times$10$\times$32\,cm$^3$) using commercial off-the-shelf components, the majority of which have space heritage. We demonstrate how to leverage the latest advancements in nano-satellite body-pointing to show that our 4\,kg CubeSat can provide performance comparable to much larger 600\,kg satellite missions. A comprehensive link budget and simulation was performed to calculate the secure key rates. We discuss design choices and trade-offs to maximize the key rate while minimizing the cost and development needed. Our detailed design and feasibility study can be readily used as a template for global scale Q.Com.
\end{abstract}




\end{abstractbox}
%

\end{frontmatter}



\section{Introduction}
The security of quantum communication (Q.Com) is based on fundamental and immutable laws of physics and not on the hope that a problem is too difficult for an adversary to solve. Naturally, this future-proof and unconditionally secure communication technology has a large impact on global communications. Attempts to overcome the limits imposed by losses, such as Ref.~\cite{Azuma2016}, and attempts to create a global satellite based network are underway~\cite{Yin2017, Liao2017a}. The latter are gigantic and incredibly complex ultra-modern satellites which can cost upwards of 100 million\,\euro\,each. Small CubeSats however can be constructed and launched for 1 to 10 million\,\euro.
We used an iterative Size, Weight and Power (SWaP) optimization process to create a design for the simplest, smallest, lightest and least power-consuming satellite system capable of Q.Com with a commercially viable key rate. We studied previous long distance implementations via optical fiber~\cite{Yin2016}, free space terrestrial links~\cite{Ursin2006} and the  successful 600\,kg class~\cite{Yin2017} and 50\,kg class~\cite{Takenaka2017} large satellites. By analyzing the results of these proof-of-concept missions and evaluating their performance in both the uplink and the downlink scenario, we find that benefits due to the fractionally larger key rate during downlink are completely outweighed by the lower cost, ease of operation ahd simplicity offered by an uplink to the satellite. Additionally, an uplink allows for a larger variety of implementable Q.Com protocols (i.e., future-proof nature). This is because many different Q.Com protocols (e.g., E91~\cite{Ekert1991}, BB84~\cite{Bennett2014}, decoy state protocol (DSP)~\cite{Lo05a}, BBM~\cite{Bennett1992}, B92~\cite{Bennett1992a}) rely on nearly identical detection schemes for the receiver and can thus all be implemented on our CubeSat.
Previous studies such as Ref.~\cite{Ursin2009, Scheidl13, Bedington2017} have shown that space-based Q.Com is in principle feasible and culminated in two successful Q.Com satellites. Recent efforts have evaluated the feasibility of downlinks~\cite{Oi2017} while others have attempted to solve the technological challenges identified by space certifying detectors and sources of entanglement~\cite{Tang2016}. However, no previous works have evaluated the feasibility of Q.Com uplinks to satellites as small as a 3U CubeSat.

The CubeSat design considered here, will also measure light pollution stemming from the ground with a narrow field of view and thus establish a global map. This is crucial to finding dark areas near potential Q.Com customers and for other, more general applications. The timing resolution of the single photon detectors enables pulse-position-modulation in classical communication from ground to space, with exceptionally fast data rates.
The extremely sensitive single photon detectors can also be re-purposed for other terrestrial and astronomical observations with an exceptional cadence and narrow FOV. In this manuscript we nevertheless focus on Q.Com, since this objective drives the design for the satellite infrastructure. 
\subsection{Quantum communication protocols}\label{protocols}
\begin{figure}[htb!]
 \centering
 \includegraphics[height=7cm]{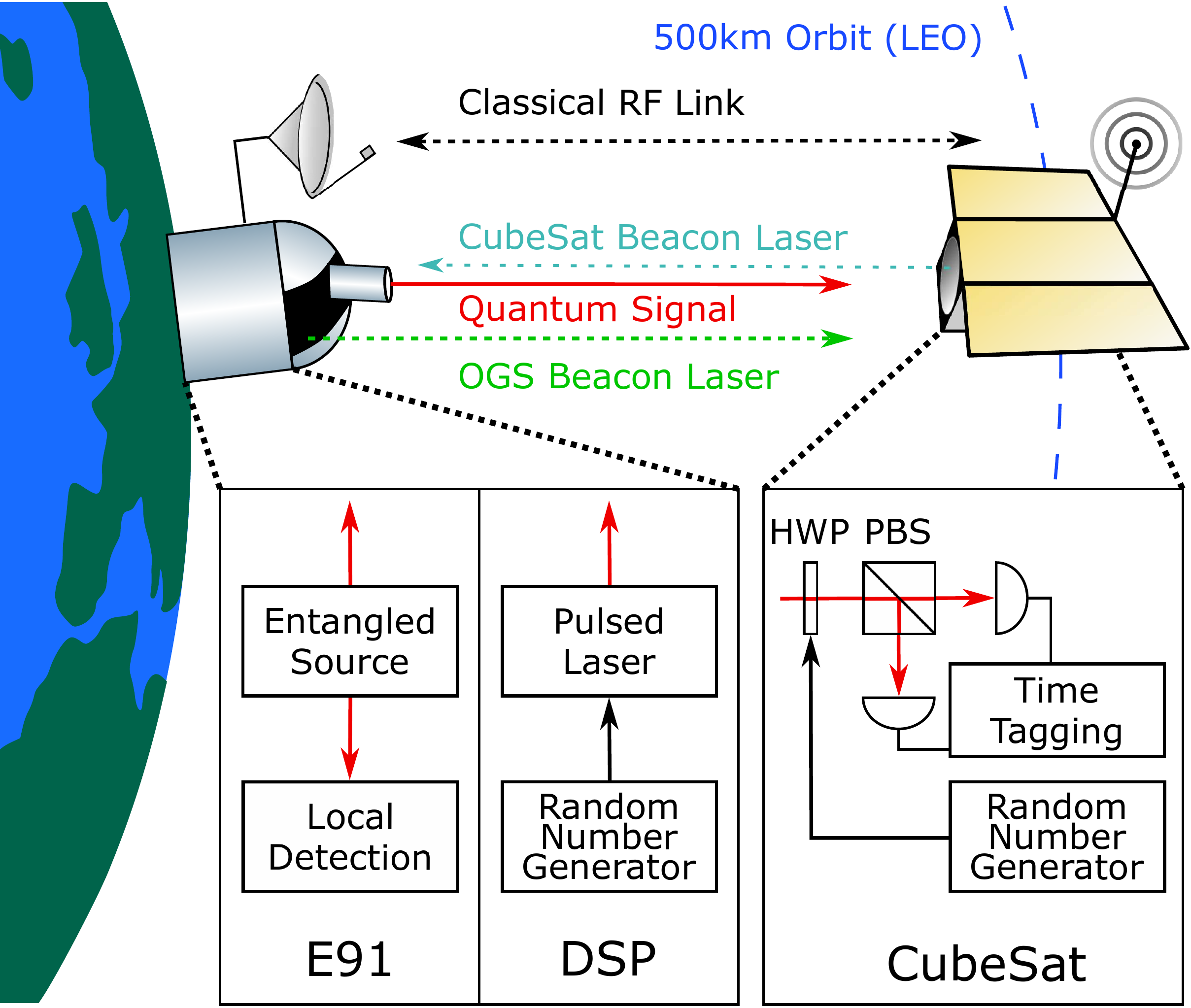}
 \caption{The optical ground station (OGS) is connected either to one arm of a source of polarization entangled photon pairs (E91) or to a pulsed laser with randomly chosen polarization and mean photon number for each pulse (DSP). The signal photons are transmitted to the CubeSat in a 500\,km low-earth orbit (LEO) via a free-space link.  OGS and CubeSat point beacon lasers at each other for precise attitude control. The quantum signal is analyzed on board the CubeSat  using a randomly switched half-wave plate (HWP) and a polarizing beam splitter (PBS). Measurement outcome, basis choice and time tag of each event are recorded. Information about the latter two is transmitted to the OGS using a classical radio frequency (RF) link. The OGS identifies the matching bits using a cross-correlation analysis ($g^{(2)}$) and tells the CubeSat which ones to use. Both disregard the other bits, perform post-processing and then share a sifted key.\label{setup}}
\end{figure}

Let us consider the two most common Q.Com protocols -- E91\,\cite{Ekert1991} and the decoy state protocol (DSP) \,\cite{Lo05a} which are explained in detail in Ref.~\cite{Gisin02, Scarani2009}.
In both, information is encoded in the polarization state of single photons at the ground station (Alice) who then sends these states to the satellite (Bob). Bob measures the polarization of the received photons in a set of randomly chosen bases. The protocol is divided into several individual ``trials''. In each trial, one state is sent and received. The techniques used to identify each trial depend on experimental implementation and protocol. To ensure that the key is secure, Alice and Bob perform statistical tests (i.e., compute the Quantum Bit Error Rate (QBER $E$)~\cite{Sergienko2005} and/or perform a Bell test) on the data they measured from several trials. Thus, they also need a form of (insecure but authenticated) classical communication. To obtain the key, Alice and Bob need various post-processing (PP) steps (detailed in~\cite{Fung2010}) that vary between protocols~\footnote{The PP steps also require a classical communication channel}. Importantly, the larger the measured QBER, the more information an eavesdropper (Eve) could, in principle, obtain about the key. Thus the number of secure bits of key that can be exchanged per second depends on the QBER.
The key difference between the protocols is that E91 exploits quantum entanglement of photons to obtain mutually shared randomness (the key) between the two parties. In DSP however, Alice encodes information by randomly choosing the polarization of an emitted weak coherent pulse. Alice must also randomly choose the average intensity of each pulse (to designate it as a signal or decoy pulse) in order to avoid a photon number splitting attack. Thus each protocol needs a different source on ground as seen in Fig.~\ref{setup} (such as Ref.~\cite{Kim2006} for E91 and Ref.~\cite{Zhao06} for DSP).
\section{Error budget}\label{errors} 
The security proofs for both E91 and DSP show that a secure key can be exchanged only if the QBER $E$ is below a certain value. For E91, the overall limit $E_{E91}^{max}$ is 11.0\,\%~\cite{Shor00}, assuming optimal classical PP with error correction efficiency $f=1$. Realistic PP techniques limit $E_{E91}^{max}$ to 10.2\%, assuming a PP efficiency of $f=1.1$~\cite{Elkouss2009}. For DSP with the same $f$ and assuming the values from Table~\ref{val}, the limit $E_{DSP}^{max}$ is about 6.2\%\footnote{Because the information entropy factor (in square brackets) depends on the gains for DSP (see Eq.~\ref{RsecDSP}), there is no constant limit for DSP, it depends on losses and average photon number per pulse. The value given is a mean value for the loss scenarios considered by us.}. These security requirements can be reformulated in terms of the more familiar Signal to Noise Ratio (SNR) as
\begin{equation}
SN\!R = \frac{1}{E}-1\,.
\end{equation}
For unconditional security, any and all noise must be attributed to Eve. This requires a minimum SNR for E91 (DSP) to 8.8 (15.1) for realistic PP with $f=1.1$. Nevertheless, we shall continue  using $E$ (instead of the SNR) to be compatible with existing literature.
Based on the formulas devised in Ref.~\cite{Ma07}, the QBER for the E91 protocol can be written as
\begin{align}
E_{E91} = &e_0- \frac{1}{Q_{E91}}\bigg[  \\ \nonumber &\frac{(e_0-e_d)\Lambda_A\Lambda_B\mu_{\scriptscriptstyle{E91}}(1+\frac{\mu_{E91}}{2})}{(1+\Lambda_A\frac{\mu_{E91}}{2})(1+\Lambda_B\frac{\mu_{E91}}{2})(1+\Lambda_A\frac{\mu_{E91}}{2}+\Lambda_B\frac{\mu_{E91}}{2}-\Lambda_A\Lambda_B\frac{\mu_{E91}}{2})} 
\bigg]\,,
\end{align}
where
\begin{align}
Q_{E91}=1-\frac{1-Y_{0B}}{(1+\Lambda_B\frac{\mu_{E91}}{2})^2}+\frac{1-Y_{0B}}{(1+\Lambda_A\frac{\mu_{E91}}{2}+\Lambda_B\frac{\mu_{E91}}{2}-\Lambda_A\Lambda_B\frac{\mu_{E91}}{2})^2}\,,
\end{align}
is the gain (or the probability of coincident photon detection per trial), $\Lambda_A$ ($\Lambda_B$) is the total loss in the channel to Alice on ground (to Bob on the satellite), $e_0$ denotes the probability of a dark count to yield an error and $e_d$ is the probability of a photon being detected in the wrong detector. The average photon number per trial is $\mu_{\scriptscriptstyle{E91}}=R^P_{E91}\tau$ (where $R^P_{E91}$ is the E91 source's pair production rate and $\tau$ is the coincidence time window). The dark count yield (or probability that a dark count occurs per trial) at the satellite is defined as $Y_{0B}=R_{B+D}\cdot\tau$ (where $R_{B+D}$ is the total rate of noise counts on the CubeSat). The effect of even several thousand noise counts on the ground based detectors is negligibly small compared to expected single count rates of $\approx\!10^7$\,cps, thus we neglect the probability of a noise count occurring at Alice ($Y_{0A}\approx0$). The secure key rate (i.e., bits per second) $R_{E91}^S$~\footnote{In information theory, ``rate'' is a normalized quantity related to entropy. However, throughout this paper we continue to use the common definition of rate as number of occurrences/instances per second.} follows directly from these quantities:
\begin{align}
R_{E91}^S\geq\frac{1}{2}\frac{Q_{E91}}{\tau}\Big[1-(1+f)H_2(\overline{E_{E91}})\Big]\,, \label{RsE91}
\end{align}
where the factor $\frac{1}{2}$ is due to the fact that only half of all basis choices are compatible, $H_2(x)$ is the binary Shannon entropy
\begin{align}
    H_2(x)=-x\log_2(x)-(1-x)\log_2(1-x)\,,
\end{align}
and $\overline{E_{E91}}$ is the QBER averaged over one measurement run where start and stop of the measurement have been chosen such that the temporal integral of $R_{E91}^S$ over one connection is maximized.
These quantities can analogously be defined for DSP, this time following Ref.~\cite{Lo05a}. The total QBER $E_{DSP}$ is given by
\begin{align}
E_{DSP}=\frac{e_d\,(1-e^{-\mu_{DSP}\cdot\Lambda_B})+e_0Y_{0B}}{Q_{DSP}}\,,
\end{align}
with the total gain $Q_{DSP}$ given by
\begin{align}
Q_{DSP}=1-e^{-\mu_{DSP}\cdot\Lambda_B}+Y_{0B}\,.
\end{align}
We choose the mean photon number per trial (or signal pulse) $\mu_{DSP}$=0.64 in order to maximize the secure key rate $R_{DSP}^S$. Unlike $\mu_{E91}$, the mean photon number per pulse in DSP $\mu_{DSP}$ can in practice be chosen arbitrarily, since the pulses originate directly from a (strongly attenuated) pulsed laser and not from inefficient spontaneous parametric down-conversion (SPDC) taking place in a nonlinear crystal. However, since for DSP only the true single-photon pulses can be assumed to contain secure bits, their individual QBER $E_{DSP}^1$ and gain $Q_{DSP}^1$ also have to be defined:
\begin{align}
E_{DSP}^1&=\frac{e_d\Lambda_B+e_0Y_{0B}}{\Lambda_B+Y_{0B}}\nonumber\\
Q_{DSP}^1&=(\Lambda_B+Y_{0B})\mu_{\scriptscriptstyle{DSP}}
e^{-\mu_{\scriptscriptstyle{DSP}}}\,.
\end{align}
The secure key rate can now be calculated as
\begin{align}
R_{DSP}^S\geq\frac{1}{4}R_{rep}\,\mu_{DSP}\Big[Q_{DSP}^1\big(1-H_2(\overline{E_{DSP}^1})\big)-f\,Q_{DSP}H_2(\overline{E_{DSP}})\Big]\,,\label{RsecDSP}
\end{align}
where the factor $\frac{1}{4}$ is due to the fact that only half of the photons are measured in the right basis and another half are non-usable decoy states. $R_{rep}$ is the repetition rate of the DSP source. Analogous to Eq.~\ref{RsE91}, $\overline{E_{DSP}^1}$ and $\overline{E_{DSP}}$ denote the QBERs averaged over one measurement run.
Using the realistic values shown in Table~\ref{val}, we can calculate the amount of loss each protocol can tolerate. The total link loss to the satellite, $\Lambda_B$, for E91 (DSP) must be better than -62.7\,dB (-61.2\,dB) in order to obtain a secure key, i.e. achieve a SNR of more than 8.8 (15.1). Accounting for losses in the apparatus of Alice and Bob, the maximum tolerable link loss $\Lambda_L$ from sending lens to receiving lens alone, is -43.6\,dB (-42.2\,dB) for E91 (DSP).
\begin{table}[htb!]
\begin{tabular}{!{\color{black}\vrule}l l r!{\color{black}\vrule}} 
 \hline
    \textbf{Symbol} & \textbf{Parameter}                                    & \textbf{Value}        \\ \hline\hline
    $d_B$           & Detector active area on CubeSat                          & 20\,\textmu m       \\ \arrayrulecolor{gray}\hline
    $D_A$           & OGS telescope diameter                                & 30\,cm                   \\ \hline
    $D_B$           & CubeSat telescope diameter                            & 10\,cm                 \\ \hline
    $e_0$       & Probability of noise count to be correct  & 50\%                   \\ \hline
    $e_d$       & Probability of erroneous detection         & 2\%                   \\ \hline
    $E^{max}_{E91/DSP}$      & Maximum tolerable QBER for E91 / DSP        & 10.2\% / 6.2\%                   \\ \hline
    $\eta_A$        & OGS multiplexed SNSPD efficiency (E91 only)             & 70\% ($-$1.5\,dB)       \\ \hline
    $\eta_B$        & CubeSat detector efficiency                           & 15\% ($-$8.2\,dB)       \\ \hline
    $f$           & Error correction protocol efficiency          & 1.1               \\ \hline
    $f_B$           & Effective focal length CubeSat telescope                        & 40\,cm                 \\ \hline
    $f_{SY\!N}$       & Repetition rate of OGS's beacon laser            & 10\,MHz                        \\ \hline
    FOV             & Field of view CubeSat (full angle)                    & 50\,\textmu rad      \\ \hline
    $\lambda$       & Signal photon wavelength                              & 810\,nm                \\ \arrayrulecolor{black}\hline\hline
    $\Lambda$       & Total loss                                            & $-$62.7\,dB (max)       \\ \arrayrulecolor{gray}\hline
    $\Lambda_A$     & Total loss OGS arm (source to detector) (E91 only)                         & 60\% ($-$2.3\,dB)       \\ \hline
    $\Lambda_H$     & Heralding efficiency (E91 only)                       & 85\% ($-$0.7\,dB)       \\ \hline
    $\Lambda_{TA}$  & OGS telescope loss (only E91)                             & $-$1.0\,dB                 \\ \hline
    $\Lambda_{TB}$  & CubeSat telescope loss                                & $-$1.5\,dB              \\ \hline
    $\Lambda_{OB}$  & CubeSat optical elements loss                         & $-$1.0\,dB              \\ \hline
    $\Lambda_{PB}$  & CubeSat pointing loss                         & $-$2.5\,dB              \\ \hline
    $\Lambda_{SB}$  & CubeSat basis switch loss                             & $-$0.5\,dB              \\ \hline
    $\Lambda_{SY\!N}$ & Loss due to errors in clock sync.              & $-$0.5\,dB                    \\ \arrayrulecolor{black}\hline\hline
    $\mu_{DSP}$           & Mean photon number per signal pulse (DSP only)             & 0.64             \\ \arrayrulecolor{gray}\hline
     $\mu_{E91}$           & Mean photon number per coincidence window (E91 only)             & 0.01             \\ \hline
    $r_0$           & Fried parameter                                       & 5\,cm - 40\,cm                     \\ \hline
    $R_A$           & OGS  count rate (E91 only)                            & 60\,Mcps             \\ \hline
    $R_B$           & CubeSat count rate (including noise)                     & 3\,kcps (max)          \\ \hline
    $R_B^{max}$     & CubeSat detectors' maximum count rate                 & 100\,kHz \\ \hline
    $R_{BG}$        & CubeSat background counts (total)                          & 80\,-\,180\,cps \\ \hline
    $R_{DC}$        & CubeSat dark count rate (per detector)                & 200\,cps                \\ \hline
    $R_{B+D}$        & CubeSat total noise counts (per detector)                & 240\,-\,290\,cps                \\ \hline  
    $R^P_{DSP}$           & Effective signal photon rate (DSP only)       & 315\,Mcps     \\ \hline
    $R^P_{E91}$           & Pair rate of entangled photon source (E91 only)       & 100\,Mcps     \\ \hline
    $R_{rep}$       & Repetition rate of single photon source (DSP only)    & 1\,GHz     \\ \hline
    $\sigma_A$      & OGS pointing precision (rms, full angle)               & 2.4\,\textmu rad\\ \hline
    $\sigma_B$      & CubeSat pointing precision (rms, full angle)           & 40\,\textmu rad\\ \arrayrulecolor{black}\hline \hline
    $t_A$           & Combined OGS detectors + time tagging jitter           & 16\,ps    \\ \hline
    $t_B$           & CubeSat detector + time tagging jitter               & 37\,ps        \\ \hline
    $\tau$          & Coincidence window                                         & 80\,ps                \\ \hline
    $t_{SB}$        & CubeSat basis switching time                          & 100\,\textmu s        \\ \hline
    $t_{TT}$        & Time tagging resolution  (on board CubeSat)                        & 10\,ps        \\ \hline
    $t_{MD}$        & Measurement duration of each chunk for clock sync. & 100\,ms  \\ \hline
    $t_{QC}$        & Maximum duration of quantum connection per pass  & 220\,s        \\
 \arrayrulecolor{black}\hline
\end{tabular}
\vspace{3mm}
\caption{ \label{val} List of parameters and values for which we assigned fixed values. Justification of these values is given in Sec.~\ref{prelim}.}
\end{table}

\section{Preliminary design}\label{prelim}
\begin{figure}[htb!]
 \centering
 \includegraphics[width=.95\columnwidth]{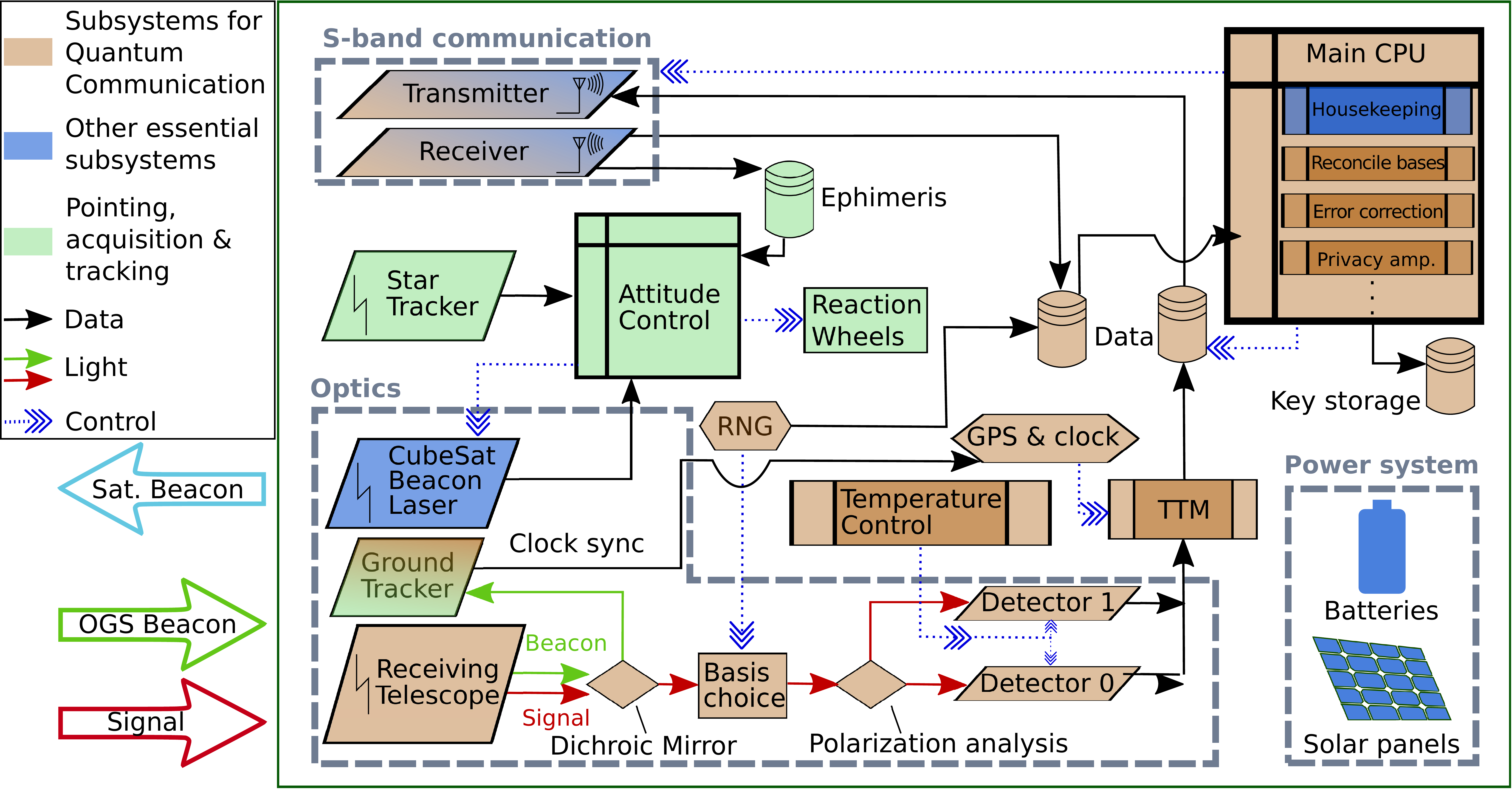}
 \caption{Block diagram of the 3U CubeSat. Components in brown are those used for quantum communication while those in green relate to the pointing, acquisition and tracking system. Other essential subsystems are blue. The subsystems can also be classified based on the type of components used as indicated by the grey dashed lines. All components are fixed to the CubeSat frame, communicate with the main CPU and supply/draw power to/from the CubeSat bus (all of which are not shown). TTM: Time Tagging Module, RNG: Random Number Generator.} \label{blockdiag}
\end{figure}

The advantage of the uplink scenario is that most of the mission's complexity is ground based and multiple protocols/experiments can be implemented. Consequently, to better understand the CubeSat design, we must first discuss the design of the optical ground station (OGS, Sec.~\ref{OGS}) and then that of the CubeSat (Sec.~\ref{cubesat}). Fig.~\ref{setup} shows an overview of the experiment consisting of space and ground segments. Table~\ref{val} provides reasonable reference values for the specifications and performance of all components as used below. Fig.~\ref{blockdiag} shows a block diagram of all payload components necessary for the Q.Com mission.
\subsection{The ground segment}\label{OGS}
Our CubeSat mission is capable of implementing a wide variety of Q.Com protocols, each of which require different sources within the OGS.
E.g. the E91 protocol requires an entangled photon source with a pair production rate $R^P_{E91}$=100\,Mcps~\cite{Steinlechner12} and an intrinsic heralding efficiency $\Lambda_H$ of 85\% ($-0.7$\,dB)~\cite{Giustina2015}. For Alice to detect these extreme count rates on ground, we suggest using multiplexed arrays of superconducting nanowire single photon detectors (SNSPDs) with a detection efficiency $\eta_A$ of 70\% (85\% for one single SNSPD without multiplexing) and a total timing jitter (including electronics) $t_A$ of 16ps (15\,ps for the SNSPD alone)~\cite{SNSPD}. This results in a total $\Lambda_A=\eta_A\cdot\Lambda_H=60$\% ($-2.3$\,dB) and a ground based detector noise rate of less than 100\,cps which we ignore in comparison to the total E91 singles rate of $R_A\approx60$\,Mcps.
Similarly, the DSP requires a source capable of producing a controllable mean photon number per pulse $\mu_{DSP} \approx$ 0.64 (0.1) for the signal (decoy) pulse where 50\% of all pulses carry a signal\footnote{Our key rate estimation based on the \emph{signal} pulse's $\mu_{DSP}$ is just an approximation without taking the photon statistics of the \emph{decoy} states into account, which have a small, but non-negligible effect on the key rate. For simplicity and in order to obtain algorithms compatible with the computing power available to us, we stick to the partial formalism outlined in Ref.~\cite{Lo05a}. For a more detailed analysis, we refer the reader to~\cite{Ma2005} and~\cite{Achilles2008}.} with a repetition rate of $>$1\,GHz. This results in an actual signal photon rate $R^P_{DSP}=315$\,Mcps at Alice. The notion of heralding efficiency $\Lambda_H$ is not applicable for DSP and can be set to 1. The same is true for imperfections in the sender optics, since any losses prior to the free-space link itself can be utilized to realize the desired $\mu_{DSP}$ value~\cite{Bourgoin13}.
All sources can be designed to produce a quantum signal at wavelength $\lambda\approx810$\,nm, which is a good compromise taking into account atmospheric absorption, Mie scattering effects, diffraction, suitable lasers for producing entanglement and suitable space based detectors (low power consumption, low dark counts and high temporal resolution). All sources also share a common sending telescope with an unobstructed diameter (to ensure a better Gaussian mode and limit the ground telescope losses $\Lambda_{TA}$ to -1.0\,dB~\cite{Sodnik}) of $D_A=30$\,cm.  The tracking precision $\sigma_A$ and slew rates of modern telescopes (typically $\sigma_A <2.4$\,\textmu rad RMS (full angle) over 5 minutes with 13$^\circ$/s slew) are an order of magnitude better than necessary to track and maintain an optical link with the CubeSat. For link calculations we assumed the OGS to be located on La Palma, where both experience from previous experiments and weather data were easily available to us. However, our design is not restricted to this location and need only be slightly adapted for areas with e.g. more cloud coverage. A suitable location for a second OGS still has to be fixed (see section~\ref{orbit}).
\subsection{The CubeSat}\label{cubesat}
\begin{figure}
\includegraphics[width=0.95\textwidth]{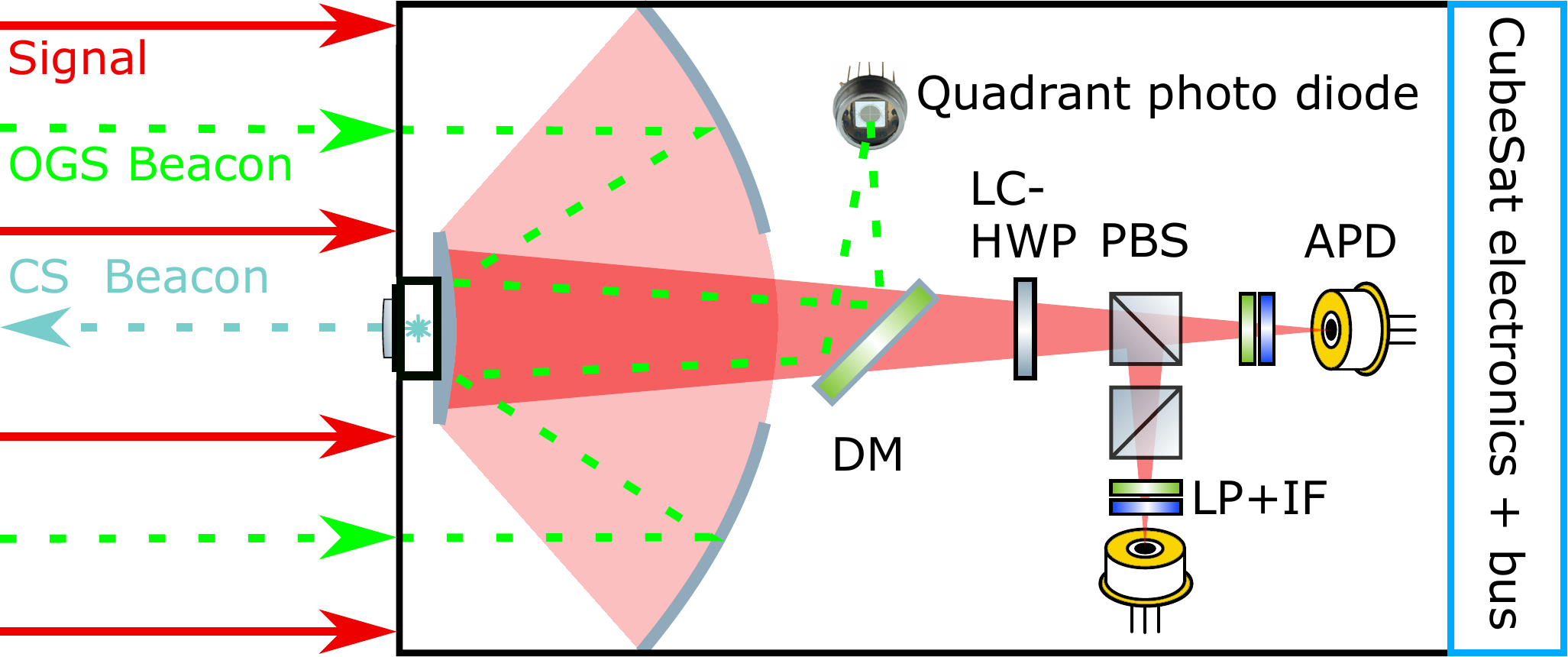}
\caption{Schematic of the optics payload on board the 3U CubeSat. The signal and beacon beams from ground are collected by a Cassegrain-type mirror telescope. The back side of the secondary mirror carries the earth-facing beacon laser necessary for tracking of the CubeSat. The input Signal and beacon are separated by a dichroic mirror (DM). The latter is tracked with a fast quadrant detector for precise attitude sensing and clock synchronization while the former passes a binary liquid-crystal-based half-wave plate switch (LC-HWP). It randomly shifts the polarization of incoming photons by either 0 or $\frac{\pi}{4}$. This effectively acts as a measurement basis switch in combination with the polarizing beam splitter (PBS) separating horizontally (vertically) polarized photons by transmitting (reflecting) them. The second PBS is used for enhanced extinction ratios. Longpass (LP) and interference filters (IF) are used to block out stray light and the photons are detected by silicon-based avalanche photo diodes (APD).}
\label{CSSetup}
\end{figure}
The CubeSat requires several subsystems as listed in Table~\ref{SWaPtbl}. Their interrelationships are shown in Fig.~\ref{blockdiag}. For a 3U CubeSat, these components must fit within 0.0032\,m$^3$, weigh less than 4\,kg
and consume less than 21Wh per orbit (with expandable $\approx 30\times30$\,cm$^2$ off-the-shelf solar panels~\cite{Solarpanels}).
Through an iterative process of SWaP design and analysis we have minimized the requirements of the CubeSat while continuing to use easy-to-obtain commercial systems. We discuss the trade-offs, design choices and compromises in Sec.~\ref{SWaP}.
Here we focus on the quantum payload which consists of receiving telescope, basis choice, polarization analysis and detection subsystems (see Fig.~\ref{CSSetup}). We estimate all optical losses within the CubeSat (between telescope and detectors) to be $\Lambda_{OB}$=$-1.0$\,dB, using only standard commercially available devices~\cite{AOS, ThorlabsIFLP, ThorlabsPBS}.

The most challenging aspect of designing a CubeSat is minimizing total noise counts $R_{B+D}$ which therefore influences many design parameters. Unavoidable stray light collected by the CubeSat's receiving telescope (i.e., background counts $R_{BG}$) and the intrinsic thermal/radiation damage counts of the detectors (i.e., dark counts per detector $R_{DC}$) add up to $R_{B+D}=R_{BG}+2R_{DC}$ and significantly degrade the SNR.
$R_{DC}$, which we assume to be constant, has to be below 200\,cps per detector to achieve a reasonable SNR. Firstly, the detector noise is reduced when operating at low temperatures. $-30^\circ$C diode temperature is desirable. Fortunately, two 250\,cm\textsuperscript{2} radiators on the sun-averted sides of the CubeSat could dissipate the 0.6\,W of thermal energy required to cool both detectors. A heating resistor should be used to further regulate the temperature to within $\pm1^\circ$C. While $R_{DC}$ of such a cooled detector can be less than 5\,cps in laboratory conditions~\cite{PDM}, it is increased by damage due to energetic particles and ionizing radiation in space. This can be mitigated by using very small active detector areas $d_B$ -- the smallest commercially available ones have a $d_B$ of 20\,\textmu m, which we expect to be small enough to keep $R_{DC}$ well below the 200\,cps limit~\cite{Anisimova2017} despite a radiation damage equivalent to a 2 year mission lifetime. Using other satellite components such as high density batteries accounts for additional radiation shielding. Other procedures to further lower the dark count rate, such as annealing the diodes, could also be implemented if necessary~\cite{Lim2017}. We therefore assume a constant 200\,cps of thermal and radiation noise per detector which is, at least for the first months of operation, a conservative estimate.

$R_{BG}$ are the erroneous measurement clicks due to near-infrared noise photons originating from the ground area which are not blocked by the spectral filters. We estimate the magnitude of this effect by using measurements of earth's luminous intensity from space~\cite{Falchi2016} considering the spectral response of the Visible Infrared Imaging Radiometer Suite (VIIRS)~\cite{VIIRS} in use. More than 50\% of the European Union's land area have less than 274\,\textmu cd/m\textsuperscript{2} night sky brightness. We divide this background intensity into contributions of artifical (light pollution mainly by HPS-lamp based street lights~\cite{Lamphar2013} which undergoes absorption through the atmosphere~\cite{MODTRAN}) and natural (earthshine~\cite{Yan2015}) sources. These calculations are valid for new moon conditions. Additionally, for the most conservative estimate, we account for scattered sunlight from a full moon (brightness: 4\,000\,cd/m\textsuperscript{2}~\cite{fullmoon}) reflected from earth (mean albedo: 0.3~\cite{Stephens2015}) into to the CubeSat (we used the solar radiation spectrum). We then translate the luminous intensity into photons~\cite{Zong2016} per~second~per~m\textsuperscript{2} footprint impinging on the CubeSat telescope with aperture $D_B=10$\,cm and calculate how many of these photons would pass through our 3\,nm wide bandpass filters centered at 810\,nm. We arrive at values of 0.55\,photons\,s\textsuperscript{-1}m\textsuperscript{-2} in zenith and 0.17\,photons\,s\textsuperscript{-1}m\textsuperscript{-2} for the lowest elevations (because of the increased distance). This effect of decreasing background close to the horizon is counterbalanced by the larger footprint of the CubeSat telescope. Optical losses and detection efficiency of the CubeSat further reduce this value (see below in this section).

In total this gives us a worst-case estimate of total noise counts which we use for all orbits regardless of the moonphase: $2R_{B+D}$ varies from $\approx480$\,cps in zenith to $\approx575$\,cps at 30$^\circ$ elevation from horizon. This assumption is very conservative, specially when compared with the 350\,cps total noise counts at full moon of a similar uplink  experiment~\cite{ren2017ground}.

If the orbit height is fixed (we chose 500\,km, see Sec.~\ref{orbit}), $R_{BG}$ can only be reduced by reducing the field of view (FOV=$\,\nicefrac{d_B}{f_B}$ where $f_B$ is the CubeSat telescope's effective focal length). This has two additional benefits: A long $f_B$ improves the polarizing beam splitter's (PBS) extinction ratio since it reduces the divergence of the impinging beam within the PBS. More importantly, a small $d_B$ strongly reduces the radiation damage to the detector due to its small cross sectional area. However, the FOV must be large enough to maintain the OGS in view despite the pointing errors of the CubeSat. Until recently, the attitude control of small CubeSats was too imprecise, requiring a large FOV that would have resulted in too many background counts to make the mission possible. The latest commercially available systems based on star trackers~\cite{BlueCanyon} have a body pointing precision $\sigma_B$ of better than 40\,\textmu rad RMS (full angle)~\cite{Duncan}. The resulting pointing losses $\Lambda_{PB}$ due to this error, which are caused by an effective spot size broadening on the detectors when averaging over time, can be shown to be
\begin{align}
    \Lambda_{PB}=1-\exp\Bigg[-\frac{\frac{1}{2}\,\text{FOV}^2}{\big(\frac{2\lambda}{\pi D_B}\big)^2+\sigma_B^2}\Bigg].
\end{align}
This attitude precision allows us to limit the FOV$<50$\,\textmu rad while introducing pointing losses $\Lambda_{PB}$ of $-2.5$\,dB. These comparably high losses are outbalanced by the strongly reduced $R_{BG}$ because of the narrow FOV. Optically tracking the beacon signal holds the potential to reduce these losses.

To achieve an optimal $f_B$, a Cassegrain-type reflector is a good choice for the receiving telescope despite the increased telescope loss $\Lambda_{TB}$ due to the secondary mirror (which we estimate to be $-1.5$\,dB in total). This is because the overall design is lightweight and the required $f_B$ of 40\,cm can be realized with a 10\,cm long telescope. The telescope covers the CubeSat's quadratic Z+ surface of about $9\times9$\,cm. For simplicity, our calculations assume a circular telescope with $D_B=$10\,cm.
Another significant challenge of Q.Com with a tiny CubeSat is the random basis choice at the start of every trial. This is necessary because Eve can exploit any predictability (or similarity between consecutive trials) of the measurement bases to gain knowledge about the key. Mechanical rotation of a half-wave plate (HWP), while sufficient for a proof-of-principle demonstration, is far too slow. Larger satellites can either use a passive basis choice (i.e., a combination of a 50:50 beam splitter (BS), two PBSs, a HWP and four detectors, such as proposed for the 12U CubeSat of Ref~\cite{1711.01886}) or an extremely fast active one (e.g., rapid Pockels cells). The former requires longer focal length telescopes and twice the number of detectors including their shielding, cooling and high voltage electronics. The latter is either power hungry or waveguide-based and extremely lossy even with small pointing errors due to the necessity of coupling into the waveguide.
Our mission design overcomes the above limitations by using a relatively slow (response times $t_{SB}\approx$100\,\textmu s~\cite{LCHWP}) liquid crystal half wave plate (LC-HWP)~\cite{Chandrasekara2017} similar to those on board the Singaporean quantum CubeSat~\cite{Tang2016}. The security of the Q.Com link can be maintained by only considering the first detection event after each random basis choice and discarding the rest. This leads to the additional basis switching loss factor
\begin{equation}
\Lambda_{SB}=\frac{1-e^{-R_Bt_{SB}}}{R_Bt_{SB}}\,,
\end{equation}
where $R_B$ is the combined total count rate of the CubeSat detectors (including noise). For a very high single count rate of $R_B$=3\,kcps, $\Lambda_{SB}$ amounts to less than $-0.5$\,dB\footnote{For the sake of completeness it has to be noted that $\Lambda_{SB}$ is the only loss which also acts on intrinsic dark counts($R_{DC}$). Since the losses are not very high and we want to avoid underestimating noise counts, this effect is omitted.}, assuming that the basis change is conditional to a detection event which can simply be achieved electronically using a gate. If a slower LC-HWP (e.g. $t_{SB}$=3\,ms) is deployed, $\Lambda_{SB}$ can go up to $-8.5$\,dB. If one keeps all measured bits irregardless of some being measured in the same basis setting, there are no such losses, but measures would have be taken to improve privacy amplification, which would inevitably also reduce the total secure key length. The LC-HWP can be driven by a trusted random number generator, e.g. consisting of shot-noise limited measurements of the noise on a diode~\cite{Schindler2003}.

After passing the LC-HWP, the photons are spatially separated by a PBS, depending on their polarization. As seen in Fig.~\ref{CSSetup}, the receiving telescope focuses the beam through the polarizing optics onto the detectors. To compensate for the angle dependent extinction ratios of the PBS and ensure $e_d\leq$2\%, another polarizer (we suggest a second PBS rotated by 90$^\circ$ due to its high transmission) must be used in the reflected arm of the first PBS.

To ensure that saturation and dead time effects do not cause losses $>$0.1\,dB, we require a maximum count rate of each CubeSat detector $R_B^{max}\gg R_B$ in the order of 100\,kHz. The detectors consist of actively quenched silicon-based avalanche photo diodes (APDs) operated in Geiger mode, placed at the output ports of the PBS. Since their $d_B$ of only 20\,\textmu m poses a tiny cross sectional area for harmful radiation, only little to no radiation shielding is required, which has a positive effect on the weight budget (see Table~\ref{SWaPtbl}).

\begin{figure}[htb!]
 \centering
 \includegraphics[width=0.95\textwidth]{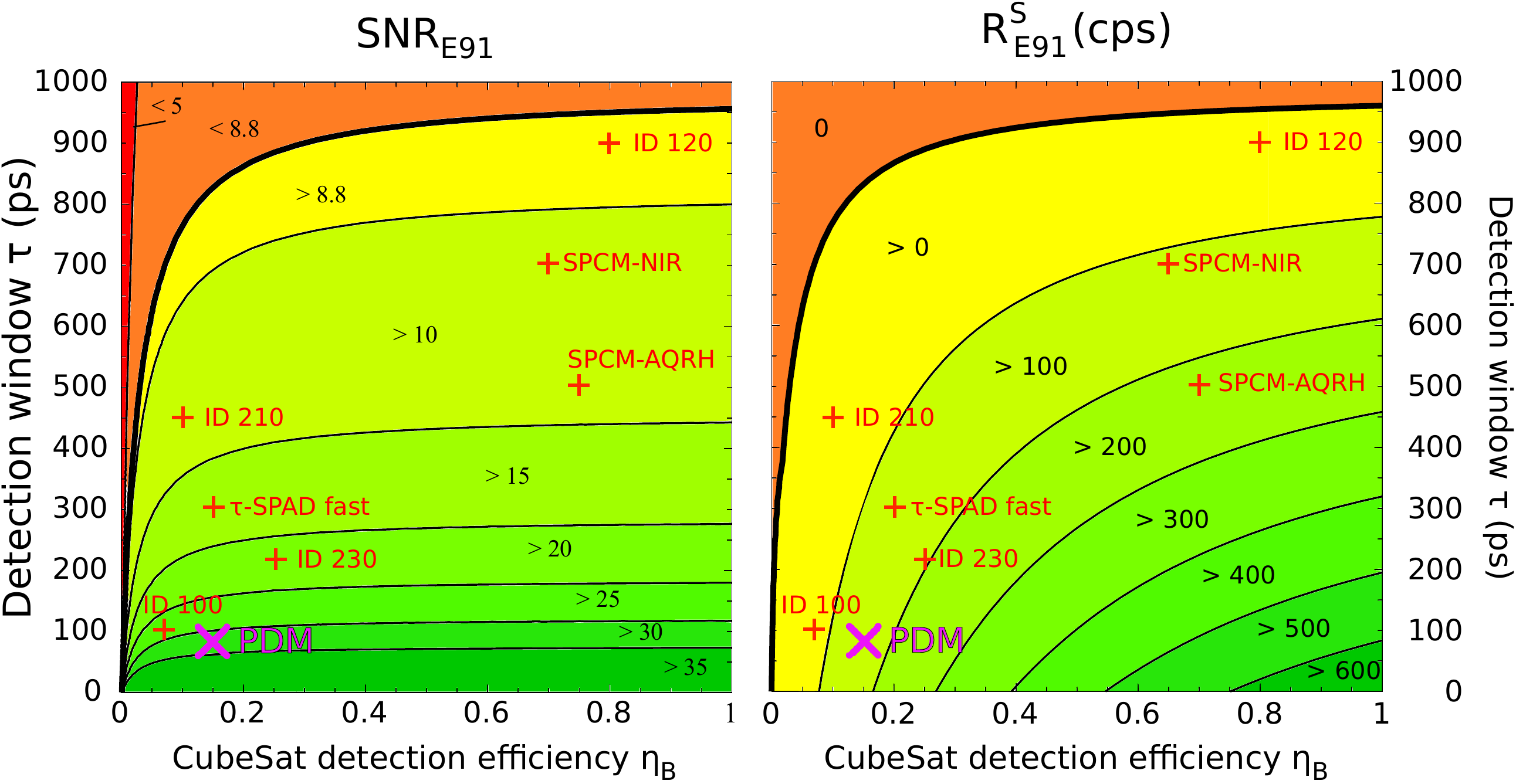}
 \caption{Contour plots of signal-to-noise ratio ($SNR_{E91}$, \textbf{left}) and secure key rate ($R^S_{E91}$, \textbf{right}) for an E91 scheme with a typical Fried parameter $r_0$=20\,cm and an elevation of $60^\circ$, showing the trade-off between CubeSat detection efficiency $\eta_B$ and timing resolution $\tau\approx2\sqrt{t_A^2+t_B^2}$ where we assue $t_A=16$\,ps to be fixed. For our study, since we want to show principal feasibility, we chose the PDM020 detector~\cite{PDM} (shown with a cross) which provides a good key rate. Other detectors are shown for comparison. 
 }\label{SNRRSEC}
\end{figure}

Errors in Q.Com depend on accidental coincidences and thus on the coincidence detection time window $\tau$. To correctly identify and distinguish at least 98\% of all pairs, $\tau > 2\sqrt{t_A^2+t_B^2}$, where $t_A$=16\,ps is the total timing jitter on ground and $t_B$ that on the CubeSat. Thus $t_B$, including the jitter of the detectors~\cite{PDM} and the time tagging electronics that note the arrival time of each pulse~\cite{TT, Swabian}, should be less than 37\,ps to ensure that we can choose $\tau$=80\,ps which is crucial to improving the SNR. The detection efficiency of the detectors we  chose is $\eta_B=$15\%. This might seem low, however we trade this for excellent temporal resolution. There is a trade-off between these two parameters: Higher detection efficiency can lead to a better secure key rate, however then the link becomes more susceptible to noise counts because of an extended coincidence detection window (see Fig. \ref{SNRRSEC}).

In addition to the quantum payload, the CubeSat optics should also accommodate an earth-facing beacon diode to aid in the ground station's tracking of the CubeSat. There should also be a dichroic mirror to separate the quantum signal from the OGS beacon. The latter assists in locating and tracking the OGS and can be detected by a  fast quadrant photo diode. The OGS's beacon signal is pulsed to facilitate clock synchronization, and the detection pulses from the fast photo diode (along with GPS signals) are used to discipline the  local clock on board the CubeSat.

It is important to consider the pointing precision of the CubeSat. Our error estimates (Sec.~\ref{errors}) show that the FOV should be $\le$50\,\textmu rad (full angle) to limit background counts. The pointing error should be of the same order of magnitude. Furthermore, to keep the OGS in view, we need slew rates of at least 1$^\circ$/s. Due to recent advances in reaction wheel and star tracker technology, commercial CubeSat attitude control can achieve a precision of 40\,\textmu rad RMS (full angle) with 10$^\circ$/s slew rate in all three axes (for a 4\,kg 3U CubeSat)~\cite{BlueCanyon}.
In addition to the transmission of photons, a fair amount of classical communication and processing is required to generate a secure key. The amount of processing done on board the CubeSat must be minimized. Thus the CubeSat will need to transmit all detection events to the OGS, which will compute coincidence events and share data identifying these sparse events with the CubeSat. Therefore the amount of data transmitted by the CubeSat far exceeds the amount of data received. We use an S-band transceiver for the actual transmission of data. Additionally, we deploy a slower UHF transceiver for housekeeping communications~\cite{SbandUHF}. Details about the data rates can be found in Sec.~\ref{sec:data} while the processing power and time required for the CubeSat to calculate the secure key is estimated in Sec.~\ref{sec:comp}.
\subsection{Preliminary SWaP analysis}\label{SWaP}
\begin{table}[htb!]
\centering
\begin{tabular}{|l|c c>{\color{red}}C{1.75cm}>{\color{red}}C{2cm}|}
\multicolumn{1}{c}{\begin{tabular}{@{}c@{}}\textbf{Subsystem name} \\ \textbf{~}\end{tabular}}&\multicolumn{1}{c}{\begin{tabular}{@{}c@{}}\textbf{Size} \\ \textbf{(U)}\end{tabular}}  & \multicolumn{1}{c}{\begin{tabular}{@{}c@{}}\textbf{Weight} \\ \textbf{(g)}\end{tabular}} & \multicolumn{1}{C{1.75cm}}{\begin{tabular}{@{}c@{}}\textbf{Peak power} \\ \textbf{(mW)}\end{tabular}} & \multicolumn{1}{C{2cm}}{\begin{tabular}{@{}c@{}}\textbf{Energy per} \\ \textbf{orbit (mWh)}\end{tabular}}\\\hline
\multicolumn{5}{|l|}{\textbf{\textit{Optics + Detection}}}\\\hline
Telescope                   & 1~~~                     & ~400  &\color{black}{-}           & \color{black}{-}           \\
Shutter                     &                       & ~100  &~5\,000        & ~~\,~~1           \\
Dichroic mirror + PBSs      & \multirow{4}{*}{0.75} & ~100  & \color{black}{-} & \color{black}{-}\\
Phase shifter               &                       & ~100  &   \multicolumn{2}{c|}{~} \\
Detectors + Shielding       &                       & ~100  & \multicolumn{2}{c|}{\multirow{2}{*}{see corresponding circuit below}} \\
Detector cooling (Peltier)  &                       & ~~50   & \multicolumn{2}{c|}{~} \\
Ground tracking photo diodes      &                       & ~100  & \multicolumn{2}{c|}{~} \\\hline \hline
\multicolumn{5}{|l|}{\textbf{Measurement control}} \\\hline
Phase shifter circuit + RNG & 0.02                  & ~~75   & ~~\,~50         & ~~\,~18 \\
Peltier circuit             & 0.01                  & ~~50   & ~1\,000       & ~~\,330 \\
Detector circuit (AQ)       & 0.07                  & ~~50   & ~~\,250        & ~~\,~46 \\
Photo diode circuit       & 0.07~                  & ~100  & ~~\,500        & ~~\,375       \\
Time tagging electronics    & 0.2                   & ~150  & 15\,000     & ~2\,750        \\\hline\hline
\multicolumn{5}{|l|}{\textbf{Positioning}} \\\hline
Beacon + electronics        & 0.01                  & ~~70   & ~1\,000      & ~~\,250        \\
XACT attitude control       & 0.5                   & ~900  & ~2\,000      & ~3\,000           \\
GPS + main computer         & 0.2                   & ~100  & ~1\,000      & ~1\,500         \\\hline \hline
\multicolumn{5}{|l|}{\textbf{RF Communication}} \\\hline
S-Band + UHF transceiver   & 0.25                  & ~114  & ~6\,000      & ~9\,000           \\
Antennas                    & 0.07                  & ~100  & ~~\,~60         & ~~\,~90        \\\hline\hline
\multicolumn{5}{|l|}{\textbf{Energy}} \\\hline
Batteries                   & 0.1~                   & ~200  & \color{green}{67\,000}     &\color{green}{60 000}          \\
Solar cells                 & -                     & ~450  & \color{green}{21\,000}     &\color{green}{21 000}   \\
Radiator                    & -                     & ~200  & \color{black}{-}          & \color{black}{-}  \\
Frame                       & -                     & ~250  & \color{black}{-}          & \color{black}{-} \\\hline\hline
\rowcolor{red}\textbf{Total consumption}  & 3.25    & 3\,759& \color{black}{31 860}    & \color{black}{17\,360}       \\\hline\hline
\rowcolor{green}\textbf{Available}    & 3.25  & 4\,000 & \color{black}{67\,000}    & \color{black}{21\,000}       \\\hline 
\end{tabular}
\vspace{3mm}
 \caption{The results of our Size, Weight and Power (SWaP) analysis along with a complete list of subsystems and their control circuits. ``Energy per orbit'' refers to consumption per one full orbit while performing a quantum measurement and takes into account different operation times for each device.}
 \label{SWaPtbl}
\end{table}

The strict limitations of SWaP consumption pose significant challenges to the satellite design. Using commercially available products, we optimized the secure key rate produced by the CubeSat while adhering to the standard restrictions. Our results are shown in Table~\ref{SWaPtbl}. All systems not described in Sec.~\ref{cubesat} are based entirely on readily available standard CubeSat components. Further customizing of certain parts would significantly lower the total SWaP consumption. The only component that would have to be modified is the time tagger, which is however within reach of current technology~\cite{TT}.

\begin{figure}[htb!]
\begin{minipage}{0.51\textwidth}
  \centering
\includegraphics[height=5.35cm]{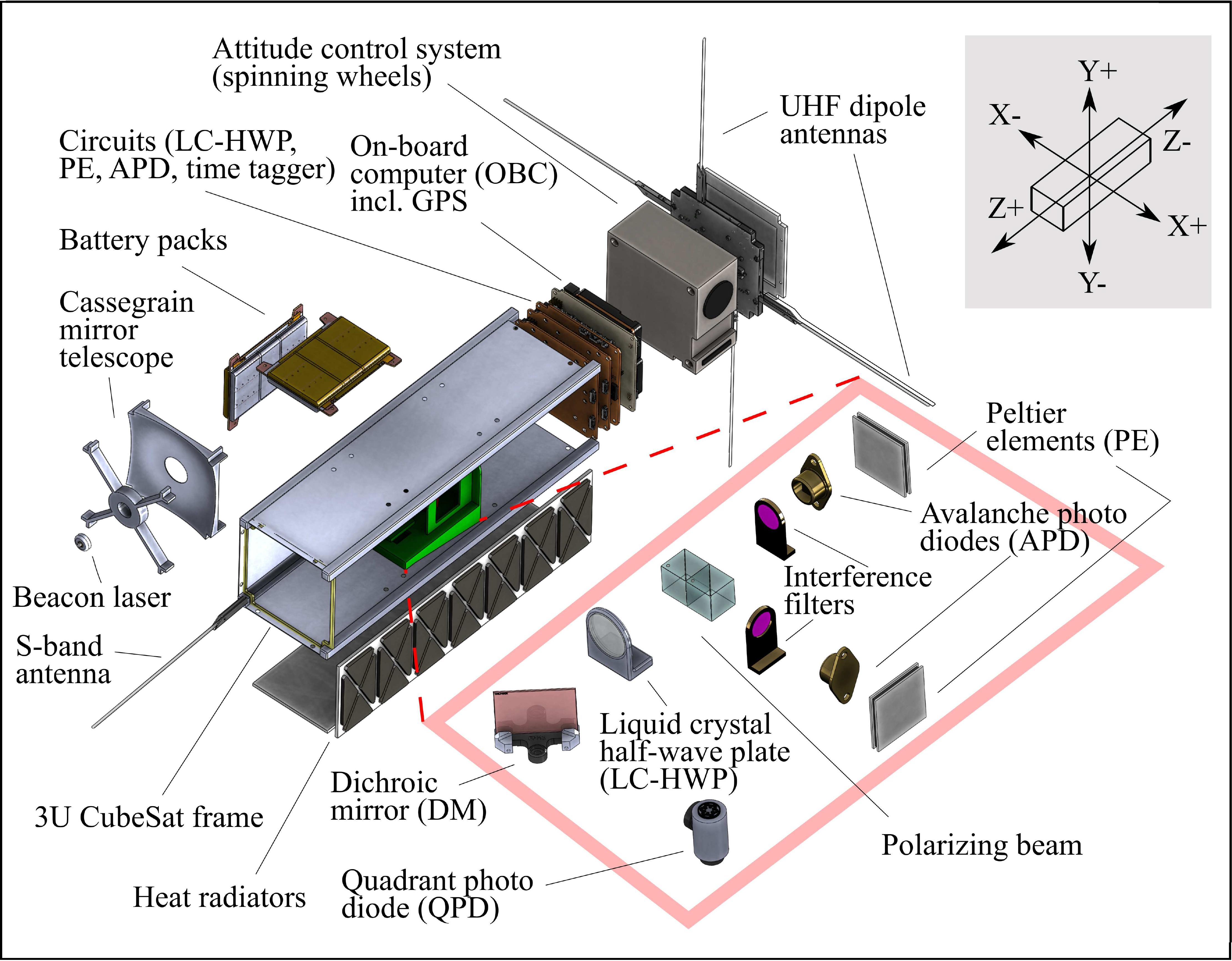}
\end{minipage}~~~~
\hfill\hfill
\begin{minipage}{0.42\textwidth}
 \raggedright
\includegraphics[height=5.35cm]{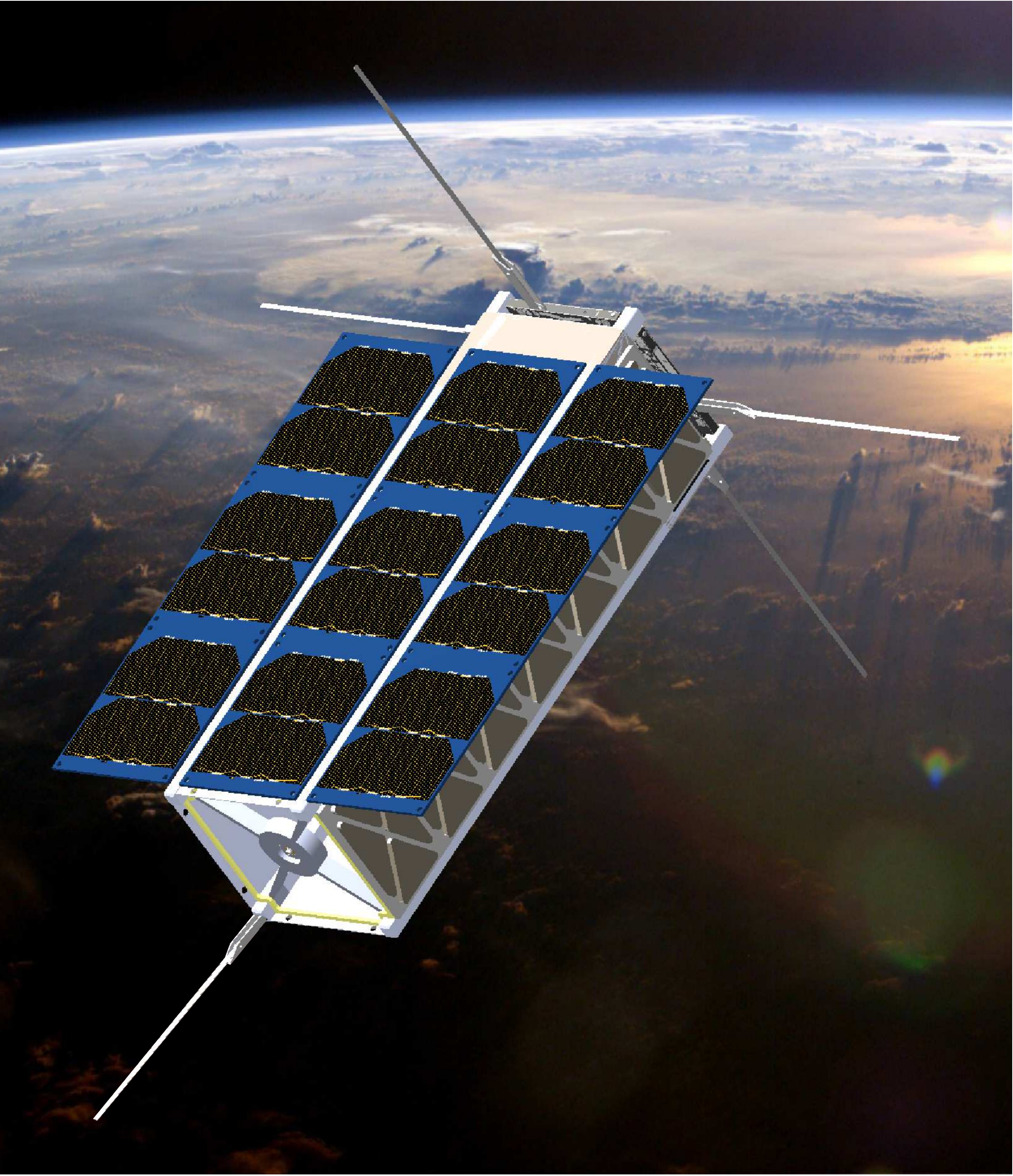}
\end{minipage}
 \caption{\textbf{Left}: Exploded view of our preliminary 3U CubeSat design. The solar panels as well as any electric connections have been omitted for clarity. The optical elements shown in the red box are out of scale. \textbf{Right}: Artistic depiction of the 3U CubeSat with expandable solar panels in bird-wing configuration. They are mounted to the sun-facing side of the CubeSat, the other three long sides of the surface can be covered with radiators for detector cooling.}
 \label{CubeSat-CAD}
\end{figure}

A standard 3U CubeSat is 10x10x34.1\,cm excluding the solar panels (with a maximum protrusion limit of 6.5\,mm)~\cite{CSspec}. We used a complete CAD model (a simplified version of which is shown in Fig.~\ref{CubeSat-CAD}) to study the actual assembly of components. Please note that we did not include size margins into our calculations since the telescope could be redesigned for a size-margin of 7\%. Also, the optics payload would allow for additional space e.g. for the batteries (as shown in Fig.~\ref{CubeSat-CAD}). However, since we restrict ourselves to off-the-shelf components, a 5\% margin is sufficient for the harness/electrical connections. Another way to gain more space would be to use the less common 4U standard (10x10x45.4\,cm)~\cite{4U}~\footnote{This would imply a launch from the ISS into an approximately 400\,km orbit.}. 

The CubeSat standard weight limit is 4\,kg for a 3U. We can include a 6\% weight margin and remain below this value. However this requirement can be relaxed to 5\,kg depending on the launch provider~\cite{spaceflight}, which is useful if an operational lifetime of more than 6 months is desired. The operational lifetime is mainly limited by radiation damage to the CubeSat, especially the APDs. Thus, heavier shielding (not included in the current SWaP) would improve operational life times at the cost of a tighter weight budget.
The type of solar panels~\cite{Solarpanels} and the orbit of the CubeSat (see Sec.~\ref{orbit}) limit the total power production per orbit to 21\,Wh. We consume only 87\% of this value.  The satellite is within line of sight of the OGS for a maximum of 11\,min (if it passes with $0^\circ$ inclination), of which at most 220\,s can be used for key generation. Thus most subsystems only operate for a fraction of each orbit. Together these consume 17.5\,Wh\footnote{The time each subsystem needs to run is calculated using conservative estimates. The detectors plus cooling are assumed to run at peak power throughout, although they only consume so much during the initial temperature stabilization phase before Q.Com starts. Similarly, the S-band data transmission is assumed to run for as long as a communication link with the OGS is possible, when in reality it need only operate for half this time.} while the always-on systems (attitude control, UHF-band communications, GPS and main computer) consume a further 13.5\,Wh per orbit.
The CubeSat must operate only at night to avoid excessive background counts. Thus a large  set of batteries are necessary. To preserve battery life and provide a safety margin we assumed that the batteries are never drained by more than 30\%. Thus we require a total battery capacity of at least 58\,Wh. Our design provides for 60\,Wh~\cite{Batteries}.
The CubeSat consumes a total of 31.9\,Wh per orbit but its solar panels can only produce 21\,Wh. This means that the CubeSat is capable of a Q.Com link roughly once every 1.5 orbits. Larger satellites would be needed for continuous operation of the Q.Com link with more than one OGS, however this drastically increases the cost. We estimate our CubeSat operating in a constellation to optimize the cost per secure key bit given the current market demand for Q.Com.

\section{Performance analysis}
{
Having specified the key parameters for the design of our CubeSat, we now want to give an estimate on the amount of secret key the satellite could acquire with two sufficiently separated OGS (one located on La Palma, the other one e.g. in Australia) over one year (Section~\ref{expsecr}). To this end, we derive a model for geometric losses due to beam divergence (Section~\ref{lossmodel}) while incorporating long-time measurements of atmospheric turbulence and weather influences (Section~\ref{weather}) to calculate different loss scenarios for our uplink. We also carry out an orbit assessment (Section~\ref{orbit}). Lastly, we evaluate the requirements for an on board clock (Section~\ref{clock}) and estimate the data storage and \mbox{-transmission} needs (Section~\ref{data}), as well as the computational requirements of the CubeSat (Section~\ref{sec:comp}).
\subsection{Optical loss model}\label{lossmodel}
The total loss $\Lambda$ consists of several contributions:
\begin{align}
    \Lambda=\Lambda_A\cdot\Lambda_B=\eta_A\cdot\Lambda_H^2\cdot\Lambda_{TA}\cdot\Lambda_{L}\cdot\Lambda_{PB}\cdot\Lambda_{TB}\cdot\Lambda_{OB}\cdot\Lambda_{SB}\cdot\Lambda_{SY\!N}\cdot\eta_B\,,
\end{align}
where $\eta_A\cdot\Lambda_H^2\cdot\Lambda_{TA}=1$ for DSP and $\Lambda_{L}$ is the link transmission from sender to receiver lens which we want to assess in this section. For a detailed justification of the values in use (which are listed in Table~\ref{val}), see Sec.~\ref{prelim} of this manuscript.
Assuming Gaussian optics, the link loss $\Lambda_{L}$ can be estimated as
\begin{equation}
\Lambda_{L}(\varphi)=\Bigg[1-\exp\Big[-\frac{1}{2}\Bigg(\frac{D_B}{w_{LP}(\varphi)}\Bigg)^2\Big]\Bigg]\cdot\Lambda_{ATM} (\varphi)\,, \label{linkloss}
\end{equation}
where $w_{LP}(\varphi)$ is the effective beam waist of the uplink signal at the satellite, depending on the zenith angle $\varphi$:
\begin{equation}
w_{LP}(\varphi)=\sqrt{w_L^2(\varphi)+\Big(\sigma_A\cdot L(\varphi)\Big)^2}\,.\label{wlp}
\end{equation}
Here, we assumed that the OGS's pointing error $\sigma_A$ follows a normal distribution, effectively increasing the divergence of the up-going beam. This is equivalent to the effect of an OGS pointing loss $\Lambda_{PA}$. $L(\varphi)$ is the distance OGS-satellite. $w_L(\varphi)$ is the beam waist at the CubeSat prior to pointing errors:\footnote{The divisor 0.316 results from the fact that any aperture passed by a real beam results in an Airy disk pattern. We consider only the innermost disk since all others' divergence is too great to hit the satellite. Now $0.316 D_A$ is the beam waist which a ideal Gaussian beam of the same intensity distribution as the innermost airy disk would have at the sending aperture, which allows us in good approximation to stick to Gaussian optics instead of having to apply Bessel functions.}
\begin{equation}
w_L(\varphi)=L(\varphi)\frac{\lambda}{0.316 D_A\pi}\Bigg[1+0.83\cdot\sec(\varphi)\Bigg(\frac{D_A}{r_0}\Bigg)^{\nicefrac{5}{3}}\Bigg]^{\nicefrac{3}{5}}\,, \label{wl}
\end{equation}
where $\lambda$=810\,nm is the photon wavelength (see Sec.~\ref{OGS}) and $r_0$ is the Fried parameter in zenith. The atmospheric attenuation $\Lambda_{ATM}(\varphi)$ in Eq.~\ref{linkloss} is given by
\begin{equation}
\Lambda_{ATM}=\exp[-\beta\cdot\sec(\varphi)]\,,
\end{equation}
where $\beta=0.22$ is the extinction optical thickness at sea level for 800\,nm~\cite{Elterman1968}.
\subsection{Weather considerations}\label{weather}
Weather conditions are crucial especially for optical uplinks since atmospheric disturbance happens immediately after the sending aperture. The Fried parameter $r_0$ gives an estimate of the atmosphere's coherence length and directly influences the upgoing beam's divergence, similar to an optical aperture. Measurements of the RoboDIMM on La Palma~\cite{RoboDIMM} over 9 years allow us to estimate the atmospheric link quality for an OGS stationed there (see Fig.~\ref{RoboDI}). Optical loss estimates for different Fried parameters can be found in Sec.~\ref{expsecr}, assuming a 500\,km orbit and 0$^\circ$ inclination with regard to the OGS. Deployment of adaptive optics systems on ground correcting for atmospheric turbulence could further decrease $\Lambda_{L}$.
\begin{figure}[htb!]
 \centering
 \includegraphics[width=0.7\textwidth]{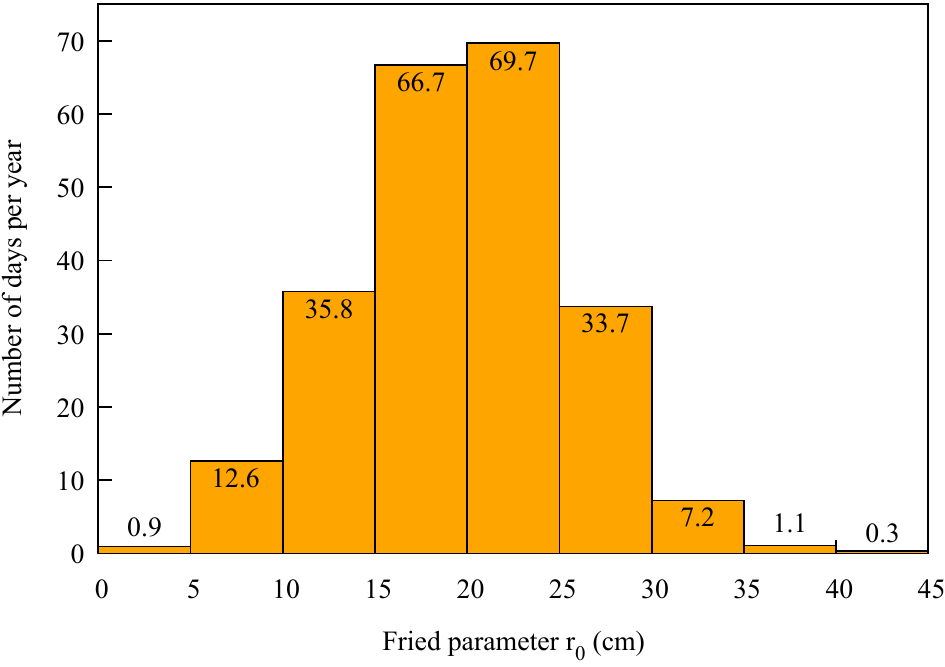}
 \caption{Histogram of days per year with certain Fried parameters $r_0$, averaged over nine years starting in February 2008. Insufficient weather conditions as well as technical problems lead to N=228 instead of 365. The average daily $r_0$ is 19.7\,cm.}\label{RoboDI}
\end{figure}
\subsection{Orbit considerations}\label{orbit}
\begin{figure}
    \centering
    \includegraphics[width=0.7\textwidth]{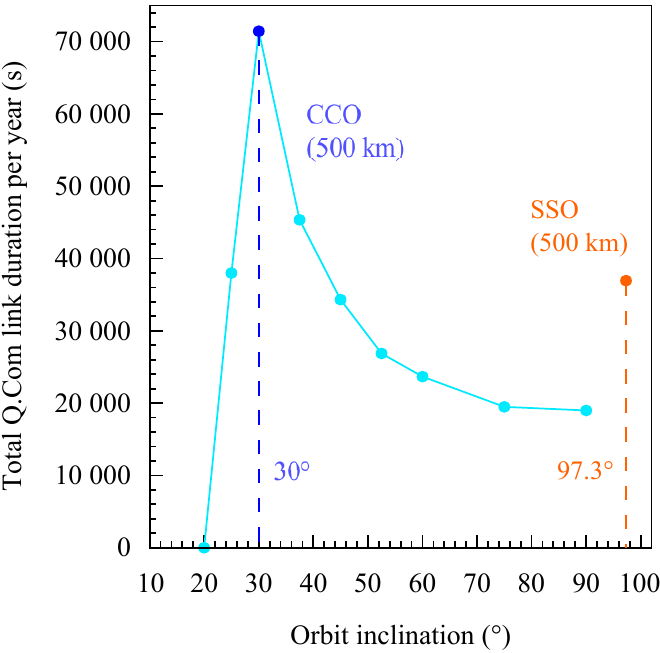}
    \caption{Variation of the total link time (during one year) for different 500\,km circular orbits (CO) and the only possible 500\,km sun-synchronous orbit (SSO) as a function of their inclination. 
    We assume that the OGS is located at  28$^\circ$\,45'\,25"\,N, 17$^\circ$\,53'\,33"\,W.
    Only passes with more than $30^\circ$ maximum elevation are considered to be contributing to the total link time.}
    \label{orbitincl}
\end{figure}
\begin{figure}
    \centering
    \includegraphics[width=0.95\textwidth]{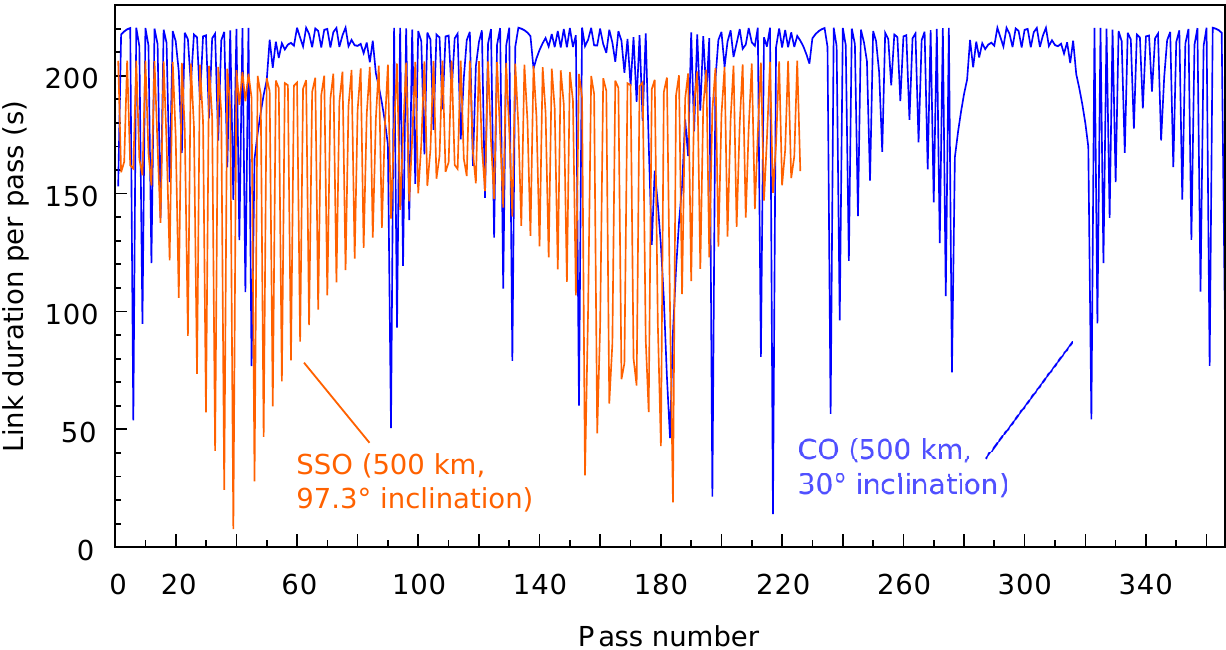}
    \caption{All orbits in both graphs are 500\,km LEO orbits that appear with more than $30^\circ$ elevation from horizon and are visible at night over La Palma.  Comparison of the link duration between two 500\,km low-earth orbits (LEOs): $97.3^\circ$ sun-synchronous (Red) and the optimal $30^\circ$ circular orbit (Blue). Only passes with more than $30^\circ$ maximum elevation at night are shown. The average visible time per pass is 195\,s (163\,s) for CO (SSO), the total visible time per year amounts to 71\,435\,s (37\,115\,s) for CO (SSO).}
    \label{orbittime}
\end{figure}
A preliminary assessment of possible orbits has to consider several limitations.  To minimize space debris, a CubeSat cannot exceed an orbit around 650-700\,km which limits the lifetime in orbit to less than 25 years~\cite{Oltrogge2011}. On the other hand, the mission shall have an operational lifetime of at least one year. This results in a minimum orbit altitude of around 400\,km (which is approximately the height of the  the ISS orbit). Subsequently, within the range between 400 and 700\,km, the choice of orbit altitude is driven by the desire to maximize the link time, relax requirements to the attitude control system as well as alignment considerations and the power budget. Considering the need to conduct the experiments during eclipse times and the fact that a significant amount of (classical) data needs to be sent between OGS and CubeSat, also orbit inclination and right ascension of the ascending node (RAAN) need to be considered. Variation of any of those orbital parameters has significant impact on the amount and duration of passes per day. We arrived at a preliminary orbit height of 500\,km (LEO). The type of orbit has an equally significant impact. An initial assessment was done to compare the link budget between a 97.3$^\circ$ inclined sun-synchronous (SSO) and a circular orbit (CO) with 30$^\circ$ inclination. For the calculations shown in  Fig.~\ref{orbittime}, a ground station on La Palma was assumed. Also note that we only considered orbital passes which allow for an actual quantum link, i.e. with elevation of $>30^\circ$ from horizon and where the OGS is in eclipse. For transmission of classical data via the S-band link, all visible orbits can be used, which amount to another $\approx40\,000$\,s ($\approx200\,000$\,s) for SSO (CO). The results shown in Fig.~\ref{orbittime} are for a mission time of one year (June 2020 to June 2021).
While the total link time of the SSO is only 37\,115\,s, the $30^\circ$ CO offers a significantly higher total link time of 71\,435\,s. Fig.~\ref{orbitincl} shows the main reason for this difference. The number of passes for a CO is significantly higher than for a SSO during one year (366 vs. 227). Also, the CO passes have a higher average link time (195\,s vs. 163\,s). The inclination of a CO and its altitude have a large impact on the link time. Fig.~\ref{orbitincl} shows the variation of link time with the orbit inclination. The best results in terms of total link time are achieved with an inclination close to the latitude of the ground station, in our case assumed to be on La Palma. 

The trapped proton flux in LEO is a significant source of radiation that can cause damage to the detectors. This radiation is significantly lower for a CO than a SSO~\cite{ginet2007energetic}.

For Q.Com between two locations on the ground, using the CubeSat as a trusted node, a second OGS would be necessary. It should be situated along the path of the CubeSat. Currently, daytime Q.Com is not possible with our scheme. However, an OGS in e.g. Australia would be able to communicate with the CubeSat during daytime in La Palma (assuming the choice of a 30$^\circ$ CO). 
\subsection{Clock synchronization}\label{clock}
Both the OGS (Alice) and the CubeSat (Bob) measure the arrival time of photons according to their own local clocks (oscillators). Nevertheless, to identify photon pairs, we must synchronize these two clocks. The precision of this clock synchronization along with the timing jitter of the detectors and electronics determines the coincidence window. Improper synchronization leads to otherwise avoidable losses.

Synchronization can be achieved using various methods such as coarse synchronization to 10\,ns using GPS~\cite{montenbruck2008precision}, exploiting the intrinsic time correlation of entangled photon pairs~\cite{Ho2009}, or using a pulsed beacon laser~\cite{Yin2017}. Clearly, GPS is too imprecise. To exploit the time correlations of photon pairs, we must measure a cross-correlation peak in the arrival times between the OGS and CubeSat.
The smallest measurement duration where we can unambiguously identify almost every coincidence peak (with the maximum acceptable total loss calculated above) is 100\,ms~\footnote{Our minimum expected pair rate is about 52\,pairs/s. To be able to correctly identify a peak, we must have more coincidence events than accidentals. With 5 coincidences we can correctly identify the peak $>$95\% of the time. Hence we choose 100\,ms as the minimum chunk duration for estimating requirements.}. 
In LEO, the velocity of the CubeSat is so large that the optical path length between the OGS and satellite can change by as much as $\approx$6\,km/s. Naturally, this causes the coincidence peak to broaden significantly. Orbital predictions and measurements can be used to correct for this. However, their typical precision is about 10\,cm~\cite{kirschner2012orbit}. This still adds a few hundred picoseconds to the coincidence window needed. 

Thus we use a pulsed beacon laser on the OGS and fast photo diodes in the CubeSat to implement a phase-locked loop and make sure that the CubeSat clock oscillates at the same frequency as that of the OGS. A beacon pulsed at  a repetition rate of $f_{SY\!N}$ = 10\,MHz coupled with a fast photo diode receiver (with, say, 1\,GHz) bandwidth on the satellite can be used to synchronize the two oscillators to within 10\,ps. Additionally, turbulence in the atmosphere can account for up to 3\,mm (i.e., $\approx$ 10\,ps) of jitter in the beacon laser's arrival time~\cite{prochazka2002atmospheric}. The effects of such phase jitter  on the received  signal can be mitigated to a large extent using a technique called jitter attenuation~\cite{jitteratten}. Nevertheless let us conservatively consider a total clock synchronization jitter of 20\,ps. Using our chosen coincidence window of 80\,ps, the above results in a synchronization loss $\Lambda_{SY\!N} < $ 0.5\,dB. 
Alternatively, we could avoid this additional loss by increasing the coincidence window to accommodate the uncertainty in the clock synchronization (i.e. the coincidence window would be 100\,ps instead of the chosen 80\,ps.).


\subsection{Data Storage and transmission}\label{data}\label{sec:data}
Since the computing power on the CubeSat is limited, Bob should send the list of all his time tags and basis choices (not measurement outcomes) down to the OGS and let Alice identify the coincidences and matching bases to tell him which counts to use. To estimate the size of data packages, we assume a time tag resolution $t_{TT}$ of 10ps~\cite{TT}. To keep the data size per tag low, it is beneficial to store just the time elapsed between consecutive events on the CubeSat. The probability $\eta_{sep}$ that the temporal separation between two successive photons will not exceed a time span $t_{max}$ during a a maximum quantum connection of duration $t_{QC}=220$\,s (see Sec.~\ref{orbit}) is given by
\begin{equation}
\eta_{sep}=\Big(1-\left(1+R_Bt_{max}\right)\cdot e^{-R_Bt_{max}} \Big)^\frac{t_{QC}}{t_{max}}\,,
\end{equation}
If one aims for a probability of less than 0.1\% for an overflow to occur during one 220\,s connection (i.e., $\eta_{sep}>$99.9\%), assuming a minimum $R_B$ of 1\,kcps because of noise counts in both detectors, $t_{max}\approx$20\,ms (result obtained numerically). This is equivalent to log$_2(\frac{t_{max}}{t_{TT}})=33$ bits per time tag including information about measurement basis and outcome. Therefore in one visible pass under optimal conditions (i.e., a 0$^\circ$ inclination overpass with an $r_0$ of 40\,cm), a maximum of 17\,Mbit of data is acquired. This means that with an 250\,kbps S-band transceiver on board the satellite, the data can be sent down in labout 70\,s~\footnote{These data transfer calculations assume an error free S-band link. Additional time or bandwidth will be needed to avoid garbled data.}. This is possible still during the Q.Com orbit if the classical transfer can be started right after the quantum link is established. Otherwise, another ground station in the satellite's path could be used or simply the next visible orbit. After Alice has calculated the correlation function and compatible basis choices, she needs to tell Bob which bits to use. Re-sending the time tags of the correct outcomes amounts to a total 3.2\,Mbit and requires another orbit since Alice has to calculate the $g^{(2)}$ in advance. In this second orbit, error correction and privacy amplification can also be carried out.
\subsection{On board computing requirements}\label{sec:comp}
The classical post processing required to obtain a secure key is not trivial and dictates the choice of the on board processing capabilities of the CubeSat. A detailed overview of these requirements can be found in Ref.~\cite{Gigov2013}. We also base our estimates provided below on the equations provided there.
The first step is to identify coincidence events. This is commonly done by computing a timing cross-correlation histogram which can be a computationally intensive task\footnote{The computational complexity of this task depends on the range of time delays that need to be scanned. Poor clock synchronization, low count rates and ever changing delays due to the satellite's motion increase the range of delays over which the cross correlation function must be computed.}. We recommend that the CubeSat share the timing of all its detection events with the OGS. The OGS can identify coincidence events and notify the CubeSat. This minimizes the amount of data transferred and the amount of calculations the CubeSat needs to perform. The on board processing of all the  $n_{tag}$ time tags should be less than $18n_{tag}$ operations in the worst case.
Calculating a sifted a key of length $m_{key}$ is estimated to require roughly $m_{key}$ bits of memory and  $15m_{key}$ operations to complete. Error correction requires additional memory and computational power. About 10 to 20\,MB of memory is sufficient for this when using algorithms based on low density parity check codes. 
Privacy amplification can be very memory efficient when using a linear-feedback shift-register-based matrix implementation and only requires memory equal to the sifted key length (i.e., $m_{key}$ bits). To estimate the processing power required, we must keep in mind that a lower SNR increases the amount of error correction and privacy amplification necessary. 
In the worst case we estimate that all these PP steps will require \mbox{$\approx~258$ million} operations per second to calculate the secure key in real time. This can easily be handled by a commercially available space certified on-board computer (OBC) with an ARM9 processor running at 400MHz with enough spare processing power for other satellite tasks~\cite{OBC}. Considering possible delays and interruptions in the classical communication link, we estimate that PP would require approximately 300\,MB of temporary memory. The OBC we consider can provide as much as 4\,GB SD card storage space. We note that the on board operating system, control programs, housekeeping functions etc. will require additional processing power and memory.
\subsection{Expected secure key rates}\label{expsecr}
\begin{figure}[htb!]
 \centering
 \includegraphics[width=0.95\textwidth]{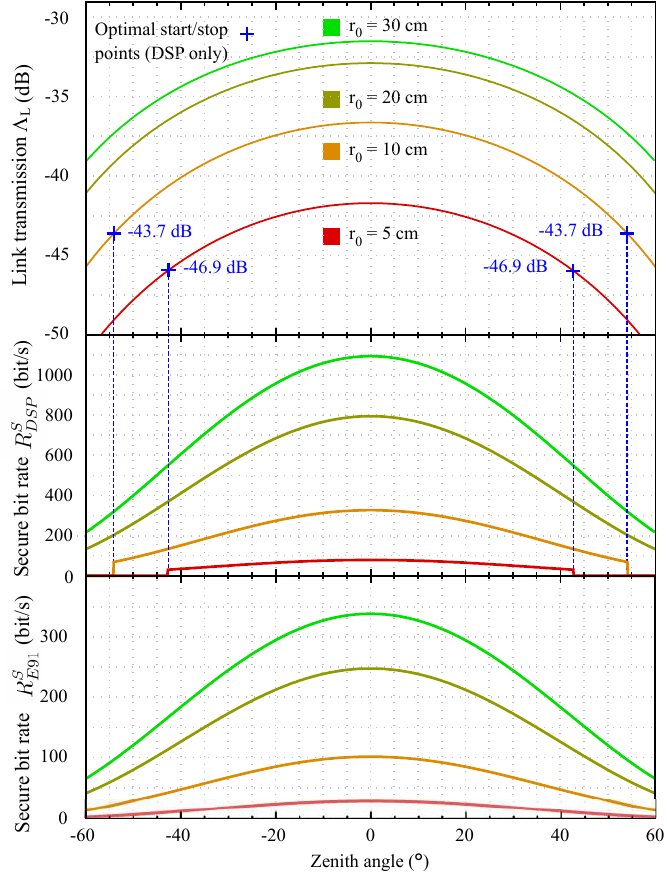}
 \caption{Link Loss $\Lambda_L$ and secure key rates $R^S_{DSP}$, $R^S_{E91}$ as functions of zenith angle $\varphi$. All curves are shown for $0^\circ$ incliniation w.r.t. the OGS. Similar curves can be calculated for different inclination angles. \textbf{Top:} Link transmission $\Lambda_L(\varphi)$ of a 500km orbit for different Fried parameters $r_0$. E91 is more tolerant w.r.t. high link losses because of its immunity to photon number splitting attacks, it can therefore always be performed during the full quantum connection. A minimum elevation from horizon of $30^\circ$ is required. \textbf{Middle:} Secure key rates for DSP using optimal starting points for values of $r_0$ that are too small to allow communication throught the orbit (marked by plusses). Other contributions to $\Lambda$ are being accounted for in these points. The temporal integral over these curves, i.e. the total key acquired during one pass, amounts to 6.8\,kbit ($r_0=5$\,cm), 33.9\,kbit (10\,cm), 95.2\,kbit (20\,cm) and 137.4\,kbit (30\,cm) respectively. \textbf{Bottom:} Secure key rates for E91 using optimal starting points for different $r_0$. The total key acquired during one pass amounts to 2.6\,kbit ($r_0=5$\,cm), 11.0\,kbit (10\,cm), 29.4\,kbit (20\,cm) and 42.0\,kbit (30\,cm) respectively.}
\end{figure}
Now that we have shown that Q.Com with a 3U CubeSat is feasible in principle, we will give an estimate of the expected key rates.  Measurements by the RoboDIMM seeing monitor on La Palma show that an $r_0$ of larger than 5\,cm can be achieved for more than 227 nights per year (see Fig.~\ref{RoboDI}) or 62\% of the time. Therefore, assuming a circular orbit with $30^\circ$ orbital inclination (see Sec.~\ref{orbit}), it can be assumed that for a total of 44\,300\,s or 12:20\,h, each year the link quality is sufficient to perform Q.Com. The average inclination in zenith as seen from the OGS, $\varphi$, is 28.3$^\circ$ (unlike the orbital pass shown in Fig.~\ref{expsecr} where $\varphi$ goes down to $0^\circ$). Computing for such an average orbit and taking the $r_0$ measurements of Sec.~\ref{weather} into account, the total key acquired in one year would therefore amount to 4.0\,Mbit (13.0\,Mbit) for E91 (DSP).
}\\

\section{Conclusion}
Q.Com offers the highest security possible. However, it is expensive and communication distances are limited. Our complete feasibility study has shown that it is possible to achieve Q.Com over thousands of kilometers using a relatively cheap and easy to construct CubeSat. By miniaturizing the design, optimizing power consumption and minimizing the weight we have shown that full-fledged commercial global Q.Com can be achieved with a simple 3U CubeSat. Not only have we shown that Q.Com is possible with such a small satellite, but we have also improved upon state-of-the-art mission designs. We have provided guidelines for building a Q.Com mission which includes selection guides for the components, trade-offs and optimizations for the secure key rate, choice of orbits etc. We discussed methods to overcome key challenges using currently available technology. We showed that the fine pointing capabilities of CubeSats no longer limit their applicability for Q.Com and optical links.

Using our CubeSat design, a pair of ground stations can exchange 13$\cdot10^6$ absolutely secure bits/year. Our CubeSat design (considering only hardware and launch) would cost $<$ 0.5 million\,\euro ~(naturally, the research/development and manpower costs for the first such satellite would be higher)~\cite{spaceflight}. Assuming a lifetime of two years, information theoretic security could be bought for as little as $\approx$60\,\euro/kbit\footnote{There are several ways to improve the cost per kbit. First, better radiation resistance and shielding would increase the lifetime of the CubeSat and proportionally decrease costs.
Second, the current cost estimate is for the interaction of one CubeSat with a pair of OGSs on opposite sides of the globe. However with careful selection, one can use multiple OGSs with the same satellite during a single orbit provided we can increase the battery capacity of the CubeSat. This would drastically reduce the costs. Third, a mass produced constellation of satellites could reduce the the cost by a further order of magnitude.}.
Key expansion protocols can be used to grow the key with only marginal security implications. Also the deployment of detectors with other characteristics can help improve the key rate at cost of the SNR (see Fig.~\ref{SNRRSEC}).
A commercially viable Q.Com satellite needs significant classical computation power, data storage and classical communication bandwidth. We have evaluated all these requirements and outlined strategies to achieve all this with minimal resources. Our future-proof planning allows for maximum versatility while utilizing minimal components on board the CubeSat. This also reduces the overall risk of the mission. Our CubeSat is compatible with the widest possible variety of polarization based Q.Com protocols. It can implement the decoy state protocol to minimize client resources or entanglement-based protocols for best verifiable security.
We have provided a complete CAD model of the CubeSat as well as a detailed discussion of the trade-offs involved in selecting components (such as those between: detection efficiency and timing jitter, radiation damage and FOV, erroneous counts and detector size, E91 and DSP, orbit of the satellite and total key etc.).
In the current design, the CubeSat is a trusted node. This is ideal for useage scenarios like communication between many branches of a single organization. The proposed CubeSat can also be used for fundamental experiments such as Bell tests which require a SNR of only 4.8 (as opposed to the SNR of 8.8 / 15.1 needed by QKD), clock synchronization, light pollution measurements and earth/atmosphere observation at the beacon wavelengths. The CubeSat can also be used to study the effect of gravity on quantum systems~\cite{Joshi2017}.

\begin{backmatter}

\section{Competing interests}
  The authors declare that they have no competing interests.

\section{Author's contributions}
 The satellite design, feasibility study and simulations were done by S.N, M.F, and S.K.J. The Mechanical CAD design was created by R.B, D.B and S.N. Orbital calculations were made by C.S. and S.A. E.K and M.B handled the systems engineering and provided valuable feedback. The effort was conceived and supervised by R.U and co-supervised by S.K.J.

\section{Acknowledgements}
  FFG Grant Nr. 4927524 / 847964, FFG Grant Nr. 6238191 / 854022, ESA/ESTEC Grant Nr. 4000112591/14/NL/US



\end{backmatter}
\end{document}